\documentclass{article}

\newcommand{\lw}[1]{\smash{\lower2.ex\hbox{#1}}}
\newcommand{\lwo}[1]{\smash{\lower1.ex\hbox{#1}}}
\newfont{\bg}{cmr10 scaled\magstep4}

\newcommand{\bigzerou}{\smash{\lower1.7ex\hbox{\bg 0}}}
\usepackage[dvips]{graphicx}
\usepackage{fancybox}

\usepackage[lflt]{floatflt}

\begin{document}
\title{\Large\bf Simulating the Effects of Quantum Error-correction
  Schemes}
\author{
{\large \hspace*{-1ex} $\mbox{\bf Jumpei Niwa}^{\ast}$ \hspace*{-
1ex}}\\
{\tt \hspace*{-2ex} niwa@is.s.u-tokyo.ac.jp \hspace*{-2ex}} 
\and
{\large \hspace*{-1ex} $\mbox{\bf Keiji Matsumoto}^{\dagger}$ \hspace*{-1ex}}\\
{\tt \hspace*{-2ex} keiji@qci.jst.go.jp \hspace*{-2ex}}
\and
{\large \hspace*{-1ex} $\mbox{\bf Hiroshi Imai}^{\ast \dagger}$ \hspace*{-1ex}}
\\
{\tt \hspace*{-2ex} imai@is.s.u-tokyo.ac.jp \hspace*{-2ex}}
}
\date{\today}
\maketitle
\footnotetext[1]{
\noindent
Department of Computer Science, Graduate School of Information Science
and Technology, The University of Tokyo, 7-3-1 Hongo, Bunkyo-ku, Tokyo
113-0033, Japan.}
\footnotetext[2]{
\noindent
Quantum Computation and Information Project, ERATO,
Japan Science and Technology Corporation,
5-28-3 Hongo, Bunkyo-ku, Tokyo 113-0033, Japan.}
\begin{abstract}
It is important to protect quantum information against decoherence and
operational errors, and quantum error-correcting (QEC) codes are the
keys to solving this problem. Of course, just the existence of codes
is not efficient. It is necessary to perform operations
fault-tolerantly on encoded states because error-correction process
(i.e., encoding, decoding, syndrome measurement and recovery) itself
induces an error. By using simulation, this paper investigates the
effects of some important QEC codes (the five qubit code, the seven qubit
code and the nine qubit code) and their fault-tolerant operations when
the error-correction process itself induces an error. The
corresponding results, statistics and analyses are presented in this
paper. 
\end{abstract}

\section{Introduction}
A quantum computer will necessarily interact with its surroundings,
which results in decoherence and consequently in the decay of the
quantum information stored in the device. Quantum gates (in contrast
to classical gates) are unitary transformations consisting of 
a continuum of possible values and hence cannot be implemented
perfectly. The effects of small imperfections in the gates will
accumulate, which will cause serious failure.

Therefore, when we build and operate a quantum computer, it is
important to protect quantum information from errors. Over the past of
few years, several theoretical studies have been made on quantum
error-correcting (QEC) codes \cite{Shor95:9bit, CS96:7bit,
  Steane96:QECC, DS95:5bit, Laf96:perfect, Got97:Phd} to protect
quantum information against errors. Of course, just the existence of
codes is not efficient. It is necessary to perform operations
fault-tolerantly on encoded states \cite{Got:FTC, Shor96:FOCS,
  Preskill97}.

The most powerful application of QEC scheme is  not only the
protection of stored or transmitted quantum information
but also the protection of quantum information as it dynamically
undergoes computation. However, it seems difficult to analyze the
effects of QEC scheme theoretically in the latter case.

For example, the main operations (i.e., computation) are interleaved
with the error-correction operations. It is necessary to prevent more
errors within a code block than the QEC code can handle. Hence, as
more errors happen in the main operations, more often the
error-correction operations need to be applied. The error-correction
operation itself, however, induces errors. Therefore, too many
error-correction operations may increase the total amount of errors.

Hence, in terms of fidelity through simulations, we investigate
the effects of QEC schemes and their fault-tolerant circuits that have
been developed by many investigators to correct errors in quantum
computations, even errors that occur during the error-correction
process. Of course, for sufficiently long computations, the
concatenated correcting codes are probably the best because of the
threshold theorem \cite{KLZ96, AB96, Got97:Phd, Preskill98}. However,
the concatenated correcting codes require much more qubits than 
simple (i.e., non-concatenated) correcting codes because concatenation
requires multi-level encodings.
Taking account of the physical implementation issue, it will be
necessary to reduce the number of required qubits as much as possible.
Therefore, we investigate how effective simple QEC codes are in the
real computation. This time we deal with the seven qubit code, the
nine qubit code and the five qubit code as the simple QEC codes
because they correct one arbitrary error but they require only less
than 10 qubits to encode 1 qubit.

Section 2 describes our simulator for quantum computing. Section 3
briefly explains the error-correction operations for the five qubit
code, the seven qubit code and the nine qubit code,
respectively. Section 4 shows the effectiveness of these QECC by the
simulations in the realistic case. Section 5 discusses related
work. Section 6 concludes by summarizing this paper.

\section{Quantum Computer Simulation System (QCSS)}
\label{sec:QCSS}
We have proposed a general-purpose fast simulator for quantum
computing \cite{NIWA02} and implemented it on various multi-computers.
The features of our quantum computer simulation system (QCSS) are
described briefly as follows.

\begin{itemize}
\item QCSS deals with quantum circuits.

  We have implemented fast libraries for fundamental operations such
  as single qubit operation, controlled operation and measurements.
  Quantum circuits can be described as a C program as shown below 
  and the fundamental operations are described as procedure calls in
  the program. The C program is compiled and linked with our libraries
  to generate executable code.

\begin{figure}[hbtp]
  \begin{center}
    \hspace*{-6cm}
    \resizebox{4cm}{!}{\rotatebox{-90}{\includegraphics*{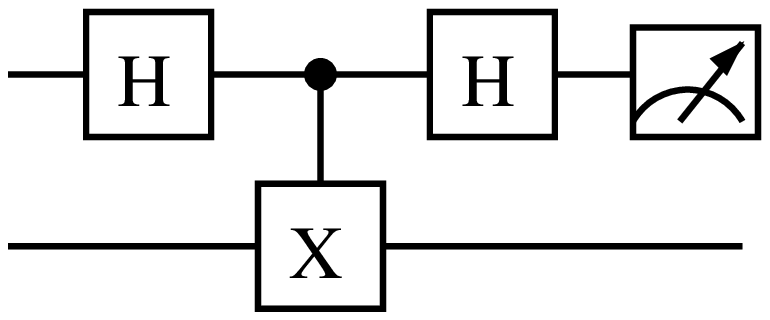}}}
    \hspace*{1cm}
    \begin{minipage}[t]{4cm}
      \begin{small}
\begin{verbatim}
ApplyGate(H, 0); /* Apply Hadamard gate to 0th qubit*/
ApplyControlledGate(X, 0, 1);/* Apply sigma X gate to 1st qubit 
                              and 0th qubit is controlled bit*/
ApplyGate(H, 0); /* Apply Hadamard gate to 0th qubit*/
Measurement(0); /* Measure 0th qubit*/
\end{verbatim}
      \end{small}
    \end{minipage}
    \caption{Sample Circuit.}
  \end{center}
\end{figure}

\item QCSS is scalable by using the parallel processing.

  QCSS is implemented on multi-computers to reduce simulation time.
  QCSS now works on not only shared-memory multi-computers (Sun,
  Enterprise4500) but also distributed-memory multi-computers
  (16-node Dell PowerEdge cluster and 128-node Sun Blade cluster).
  On the current implementation, QCSS can deal with about 30 qubits.
  Of course, this limit is caused by hardware (i.e., memory,
  processor-speed). It should be noted that QCSS can deal with more
  qubits as more powerful multi-computers are available.
  
\item QCSS implements time evolution efficiently.

  QCSS requires only ${\cal O}(2 \times 2)$ space and ${\cal O}(3
  \times 2^n)$ arithmetic operations for $n-$qubit time evolution
  \cite{Gruska99}.

\item QCSS implements the decoherence error model and operational
  error model.
  
  QCSS assumes that the quantum depolarizing channel as the
  decoherence error model. In this channel, with probability $1-p$,
  each qubit is left alone. In addition, there are equal probabilities
  $p/3$ that $X$ (a bit flip error), $Z$ (a phase flip error) or $Y$
  (both errors, $Y=XZ$) affects the qubit.

  In general, all of single qubit gates are generated from rotations
  and phase shifts. QCSS represents inaccuracies by adding small
  deviations to the angles of rotations and phase shifts. Each error
  angle is drawn from Gaussian distribution with the standard
  deviation ($\mit\sigma$).

  QCSS does not deal with mixed states. Therefore, in order to compute
  the fidelity, the experiments were repeated many times and the
  average values are used.

\end{itemize}

\section{Error-correction Circuits}

We deal with three important QECC that correct one arbitrary error. 
In this section, we introduce the error-correction operations
respectively.
\subsection{The Nine Qubit Code \cite{Shor95:9bit}}
The nine qubit code $[[9,1,3]]$ is known as Shor code.
This code is a combination of the three qubit phase flip codes and bit
flip codes. It can be seen as a two level concatenated code. The
codewords are given by:
\[|0 \rangle \mapsto |0_L \rangle \equiv
\frac{1}{\sqrt{8}}(|000\rangle+|111\rangle)(|000\rangle+|111\rangle)(|000\rangle+|111\rangle)\]
\[|1 \rangle \mapsto |1_L \rangle \equiv
\frac{1}{\sqrt{8}}(|000\rangle-|111\rangle)(|000\rangle-|111\rangle)(|000\rangle-|111\rangle).\]
\begin{figure}[hbtp]
\vspace*{-0.5cm}
  \begin{center}
    \includegraphics[width=8.5cm]{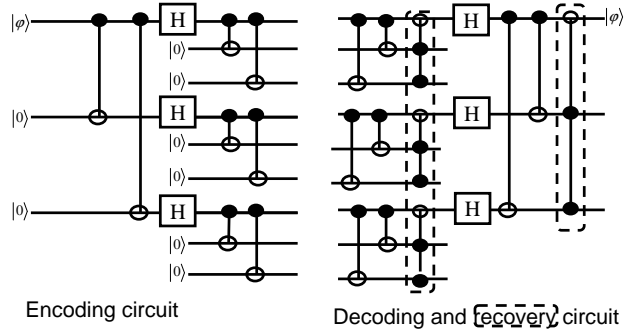}
    \caption{Encoding/decoding/recovery circuit for
    the nine qubit code}
    \label{fig:9bit}
  \end{center}
\vspace*{-1cm}
\end{figure}

Its encoder, decoder and recovery circuit are shown in Figure
\ref{fig:9bit}.
Of course, there exists the syndrome measurement and recovery circuit
based on the stabilizer formalism \cite{Got97:Phd}. However, we use
this circuit, since it is simpler and hence induces less errors. 
We should notice that this recovery circuit cannot be used without
decoding circuit.

\subsection{The Seven Qubit Code \cite{CS96:7bit, Steane96:QECC}}

\begin{floatingfigure}{7cm}
  \begin{center}
    \includegraphics*[width=6cm]{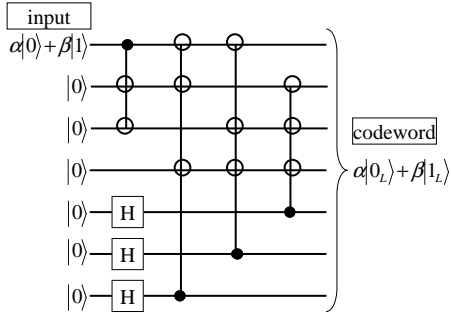}\\
    \caption{Encoder for the seven qubit code.}
    \label{fig:7e}
  \end{center}
  \vspace*{0.2cm}
\end{floatingfigure}
~~
The seven qubit code $[[7,1,3]]$ is known as
Steane code.  This code belongs to CSS code and constructed using the
$[7,4,3]$ (classical) Hamming code.  The codewords are given by: 
 $$|0_L\rangle = \frac{1}{\sqrt{8}}[|0000000\rangle +   |1010101\rangle +
  |0110011\rangle + |1100110\rangle $$
$$+ |0001111\rangle + |1011010\rangle + |0111100\rangle + |1101001\rangle$$
$$|1_L\rangle = \frac{1}{\sqrt{8}}[|11111111\rangle +
 |0101010\rangle + |1001100\rangle + |0011001\rangle$$
$$+ |1110000\rangle + |0100101\rangle + |1000011\rangle + |0010110\rangle.$$

\begin{figure}[htbp]
  \centering
  \includegraphics*[width=0.65\textwidth]{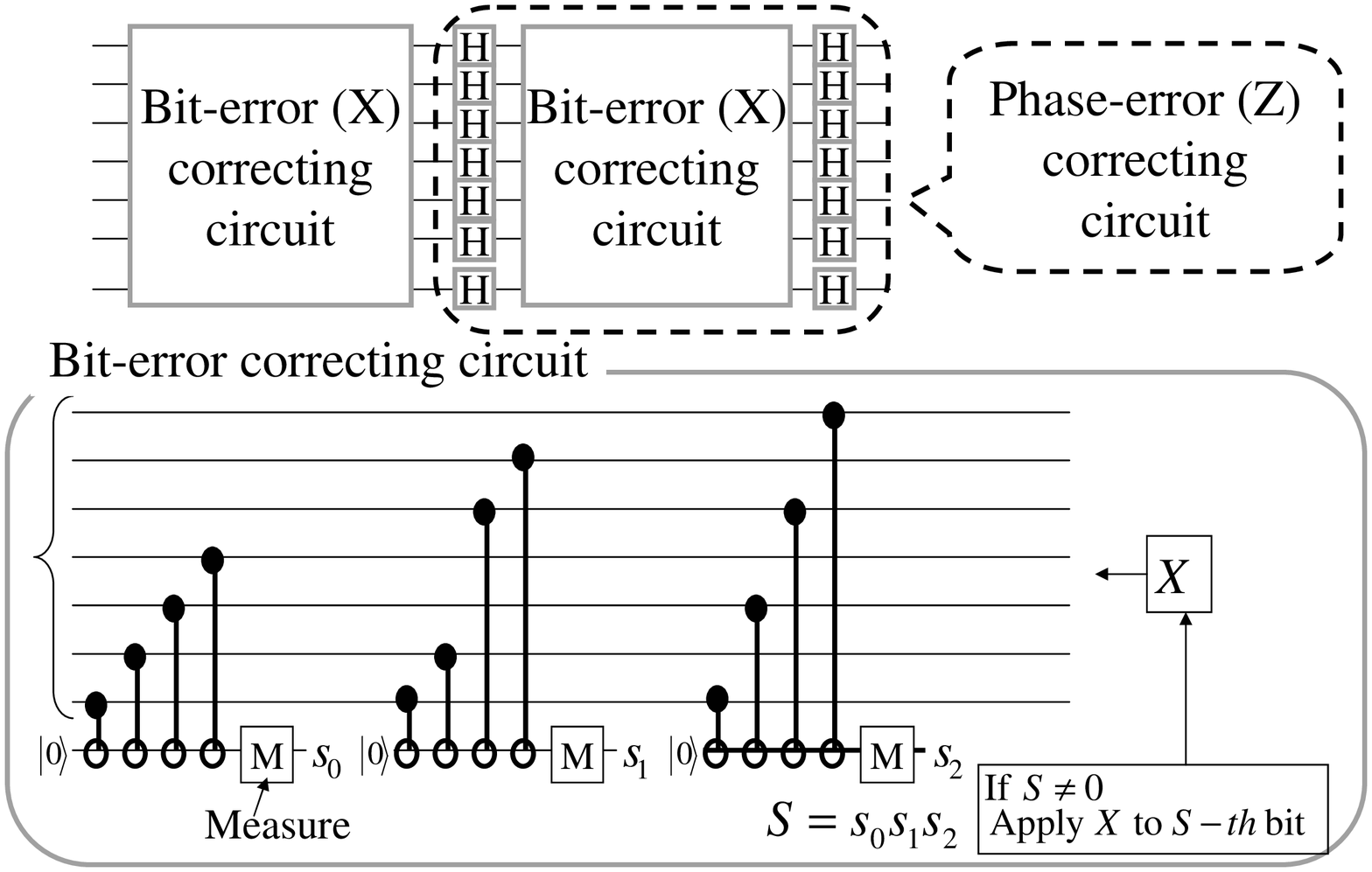}
  \caption{Syndrome measurement and recovery
  circuit for the seven qubit code.}
  \label{fig:7bitc}
\end{figure}

Steane code has nice properties. One of them is that the codewords do
no require to be decoded in order to perform main operations. Its
encoding circuit is shown in the Figure \ref{fig:7e}. Its decoding
circuit is just the encoding circuit run in reverse. Its syndrome
measurement and recovery circuit are shown in the Figure
\ref{fig:7bitc}.

\subsection{The Five Qubit Code \cite{DS95:5bit,Laf96:perfect}}
\begin{floatingfigure}{7cm}
  \begin{center}
    \includegraphics*[width=6cm]{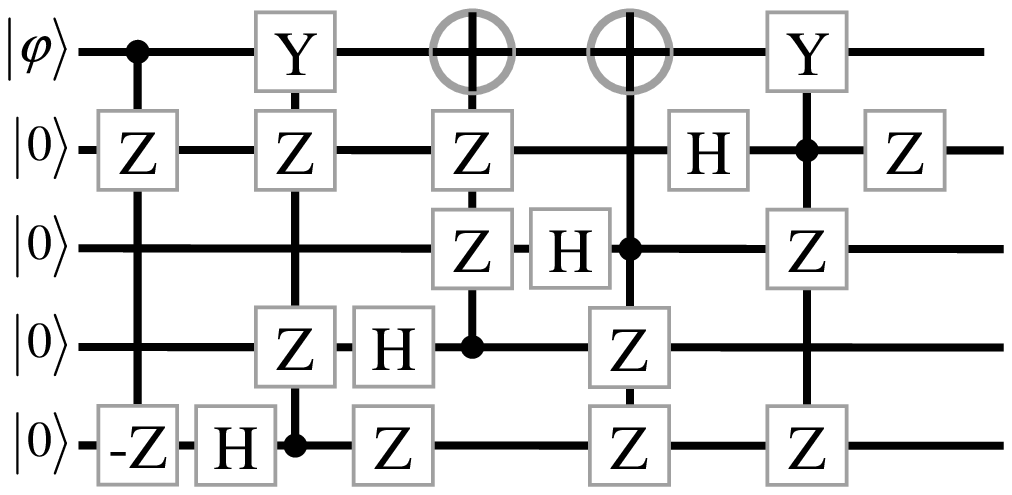}\\
    \caption{Encoder for the five qubit code.}
    \label{fig:5c}
  \end{center}
\end{floatingfigure}
~ \\

\begin{figure}[htbp]
  \centering
  \includegraphics*[width=0.72\textwidth]{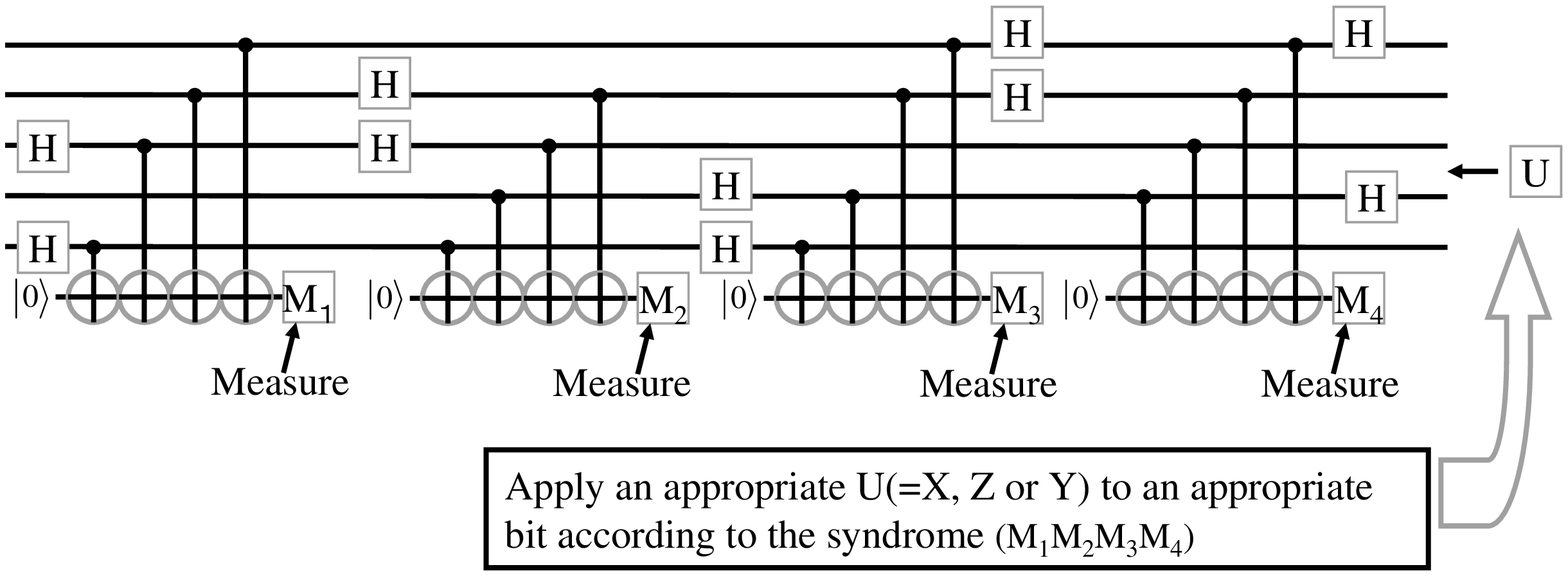}
  \caption{Syndrome measurement and recovery
    circuit for the five qubit code}
  \label{fig:5bitc}
\end{figure}

The five qubit code $[[5,1,3]]$ is the perfect non-degenerate code and
belongs to the non-CSS stabilizer code. This code is the smallest
capable of protecting against a single error.
Its encoding circuit is shown in the Figure \ref{fig:5c}. Its decoding
circuit is just the encoding circuit run in reverse. The error
syndrome measurement and recovery circuit is shown in the Figure
\ref{fig:5bitc}. The codewords are given by:

\begin{eqnarray*}
|0_L\rangle & = & \frac{1}{4}[|00000\rangle + |10010\rangle +
|01001\rangle + |10100\rangle + |01010\rangle + |00101\rangle \\
& & - |11110\rangle - |01111\rangle - |10111\rangle - |11011\rangle - |11101\rangle\\
& & - |01100\rangle - |00110\rangle - |00011\rangle - |10001\rangle -
||11000\rangle\\
|1_L\rangle & = & \frac{1}{4}[|11111\rangle + |01101\rangle + 
|10110\rangle + |01011\rangle + |10101\rangle + |11010\rangle \\
& & - |00001\rangle - |10000\rangle - |01000\rangle - |00100\rangle - |00010\rangle\\
& & - |10011\rangle - |11001\rangle - |11100\rangle - |01110\rangle - |00111\rangle]\\
\end{eqnarray*}

It will be helpful to compare the features of these three QEC codes.
Table \ref{tab:comp} summarizes the comparison among three QEC codes.
\begin{table}[htbp]
  \centering
  \caption{Comparison among the nine qubit code, the seven qubit code
    and the five qubit code}
  The number of (minimum) required qubits.\\    
  \begin{tabular}{|c||c|c|c|}
    \hline
     & $[[9,1,3]]$ & $[[7,1,3]]$ & $[[5,1,3]]$ \\
     \hline
     qubits & 9 & 8 & 6 \\
    \hline
  \end{tabular}\\
  ~\\
  Depth of error-correcting circuits.\\    
  \begin{tabular}{|c||c|c|c|}
    \hline
    & $[[9,1,3]]$ & $[[7,1,3]]$ & $[[5,1,3]]$ \\
    \hline
    Encoder (Decoder) & 5 & 4 & 10 \\ \hline
    Syn. measurement and recovery & 2 & 32 & 22\\
    \hline
  \end{tabular}\\
  ~ \\
  Transversal gate implementation \cite{Got:FTC}.\\    
  \begin{tabular}{|c|c|c|}
    \hline
    $[[9,1,3]]$ & $[[7,1,3]]$ & $[[5,1,3]]$ \\
     \hline
      Hard  & Easy  & Hard \\
    \hline
  \end{tabular}
  \label{tab:comp}
\end{table}
Table ``Depth of error-correcting circuits'' shows the depth of
error-correcting circuits using the minimum number of qubits. As for
error-correction circuit complexity, the nine qubit code is
simplest. Furthermore, it does not require an ancilla qubit. About
implementation of transversal gates, that is, encoded gates in a
bit-wise fashion, the seven qubit code is the easiest.

\section{Experimental Verification through Simulations}

The most powerful application of quantum error-correction scheme is 
not only the protection of stored or transmitted quantum information
but also the protection of quantum information as it dynamically
undergoes computation.

Quantum computation using encoded qubits, is a sequence of computation
directly (or indirectly) on the encoded qubits along with periodic
error-correction processes. If these processes are not performed
sufficiently, then multiple errors may occur between these processes,
which results in an uncorrectable state. However, if these processes
are performed too much, then the circuit size becomes too large, which
increases the total amount of errors.  Our simulation aims to
investigate how often to perform these operations.

\subsection{Simulation Methodology}
\label{sec:sm}
\begin{figure}[htbp]
  \centering
  \includegraphics*[width=0.72\textwidth]{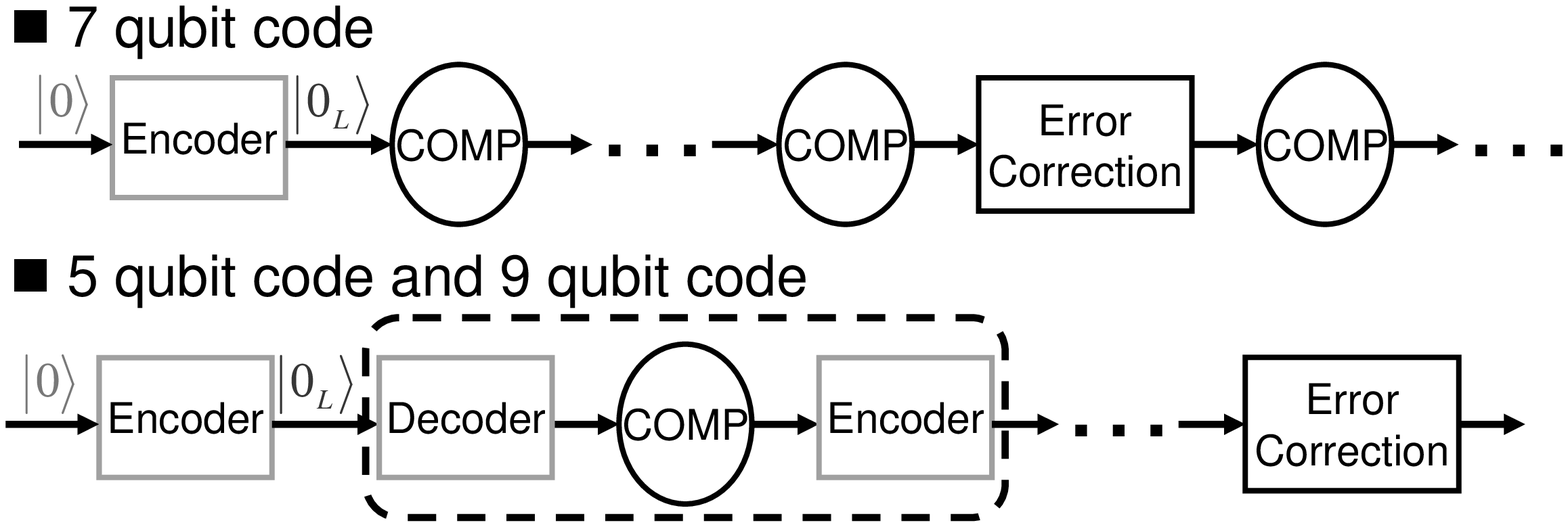}
  \caption{Simulation Methodology}
  \label{fig:sum_met}
\end{figure}

The simulation is performed using QCSS described in Section
\ref{sec:QCSS}.

As for the seven qubit code, it is easy to perform computation
directly on the encoded qubits, especially about normalizer 
operations (the Hadamard, phase and controlled-NOT gates).

On the other hand, as for the five qubit code and the nine qubit code,
it is hard to perform computation  directly on the encoded qubits
without using more additional qubits. Therefore, it is necessary to
decode codewords in order to perform main computation if we reduce the
number of qubits as much as possible. That is, encoding/decoding
operations between computation are necessary. Furthermore, as for the
nine qubit code, the recovery circuit should be used along with
decoding circuit. This means that
the error-correction operation is performed at every 1 main gate. As
for seven qubit code and five qubit code, error-correction is
performed at every 1 or 50 or 100 or 200 or 2000 main gate.

Let $p$ be the decoherence probability. At each time step, that is, as
the depth of circuit increases by 1, each qubit has no change with
probability $1-p$ and it undergoes rotations by the operation $X, Y,
Z$ with equal probabilities $\frac{p}{3}$.

QCSS represents inaccuracies by adding small deviations to the angles
of rotations and phase shifts. Each error angle is drawn from Gaussian
distribution with the standard deviation ($\mit\sigma$).

When we compute fidelity, we trace out ancilla qubits. In order to
trace out ancilla qubits, QCSS copies the state and ideally decodes
the copied state and measures only ancilla qubits. Of course, the
original state has no change in fidelity calculation.
Fidelity is calculated as follows
\[fidelity = \sum_{i=1}^{N} \frac{|| \langle \phi_{correct}|
  \phi_{output_{i}} \rangle ||^2}{N}.\]
N means the number of experiments ($10^5 \sim 10^6$).
$|\phi_{correct} \rangle$ represents the correct state 
and $|\phi_{output_{i}} \rangle$ represents the output state computed by
  QCSS at $i$-th experiment.

\subsection{Effects of Ancilla Operations}
To compute syndrome, ancilla qubits are usually used.
The obvious way is to use only one qubit in order to compute each bit
of the syndrome. In this case, ancilla preparation is very
simple. However, it may propagate the errors backward.


The fault-tolerant way is to use enough qubits to compute each bit of
the syndrome transversally. Of course, we prepare the ancilla in a
Shor state \cite{Shor96:FOCS} that reveals the errors without
revealing the data, The latter is much more fault-tolerant than the
former. However, this requires much more qubits in total and ancilla
preparation is much harder.

Thus, we need to check which system is better in terms of fidelity
through simulations.
We use the seven qubit code and adopt 4000 times Hadamard
transformation as main computation.  The start state of the quantum
register is $|0_L\rangle$ (logical 0 bit). If there are no errors, the
final correct state must be $|0_L\rangle$.
The latter error-correcting circuit requires 11 qubits in total and its
depth is at least 40 \footnote{The ancilla preparation is repeated
  until it succeeds.}. 
\begin{table}[htbp]
  \centering
  \vspace*{-0.5cm}
  \caption{Final fidelity with decoherence errors for the one qubit
    ancilla system and the four qubit ancilla system}
  \begin{tabular}{|c|c||c|c|c|c|c|} \hline
    \lw{$rate$} & {Ancilla} & \multicolumn{5}{c|}{Frequency of recovery process} \\ \cline{3-7}
    & bit & 1G & 50G & 100G & 200G & 2000G\\ \hline
    \lw{$10^{-5}$} & 1bit & 0.5840 & 0.9836 & 0.9920 & 0.9942 & 0.9925
    \\ \cline{2-7}
                   & 4bit & 0.5750 & 0.9808 & 0.9886 & 0.9955 & 0.9922 
    \\ \hline
    \lw{$10^{-4}$} & 1bit & 0.4970 & 0.8290 & 0.8910 & 0.8940 & 0.7110 
    \\ \cline{2-7}
                   & 4bit & 0.5020 & 0.8010 & 0.8558 & 0.8705 & 0.7164 
    \\ \hline

    \lw{$10^{-3}$} & 1bit & 0.4890 & 0.5070 & 0.4780 & 0.5220 & 0.5070 
    \\ \cline{2-7}
                   & 4bit & 0.4900 & 0.5082 & 0.5075 & 0.4928 & 0.5034 
    \\ \hline
  \end{tabular}
  \label{tab:ancilla}
  \vspace*{-0.2cm}
\end{table}

Table \ref{tab:ancilla} shows the final fidelity with decoherence errors
($rate=10^{-5} \sim 10^{-3}$) for the one qubit ancilla system and the
four qubit ancilla system. For example, The column ``50G''represents
the results performing the QEC circuit at every 50 main (i.e.,
Hadamard) gate. From Table \ref{tab:ancilla}, we cannot conclude that
the four qubit ancilla system is better than the one qubit ancilla
system. 

Consider the circuit area defined by ``number of the qubits'' $\times$ ``depth
of the circuit''. It is reasonable to suppose that the error
probability of the circuit is proportional to its area. The area of
the four ancilla qubits system is at least 440 and it is much larger than
that of the 1 ancilla qubit system, that is, 256. None the less, the
fidelity values are almost the same. This result indicates that the
transversal operations in the four ancilla qubits system are
fault-tolerant.

\subsection{Effects of Frequency of the Error-correction Operations}
We adopt 4000 times Hadamard transformation as main computation.  
The start state of the quantum register is $|0_L\rangle$ (logical 0
bit). Fidelity is calculated at every two main gate, which means that
$|\phi_{correct}\rangle$ is $|0_L\rangle$ if there are no errors.
From now on, the x-axis in the Figure  shows the even number of main
computations (i.e., Hadamard Transform). The y-axis in the Figure
shows the fidelity. 

\subsubsection{Decoherence Errors}
\label{sec:dec_exp}

\paragraph{The Seven Qubit Code}~\\
\begin{figure}[htb]
  \vspace*{-0.5em}
  \resizebox{0.5\textwidth}{!}{\rotatebox{-90}{\includegraphics*{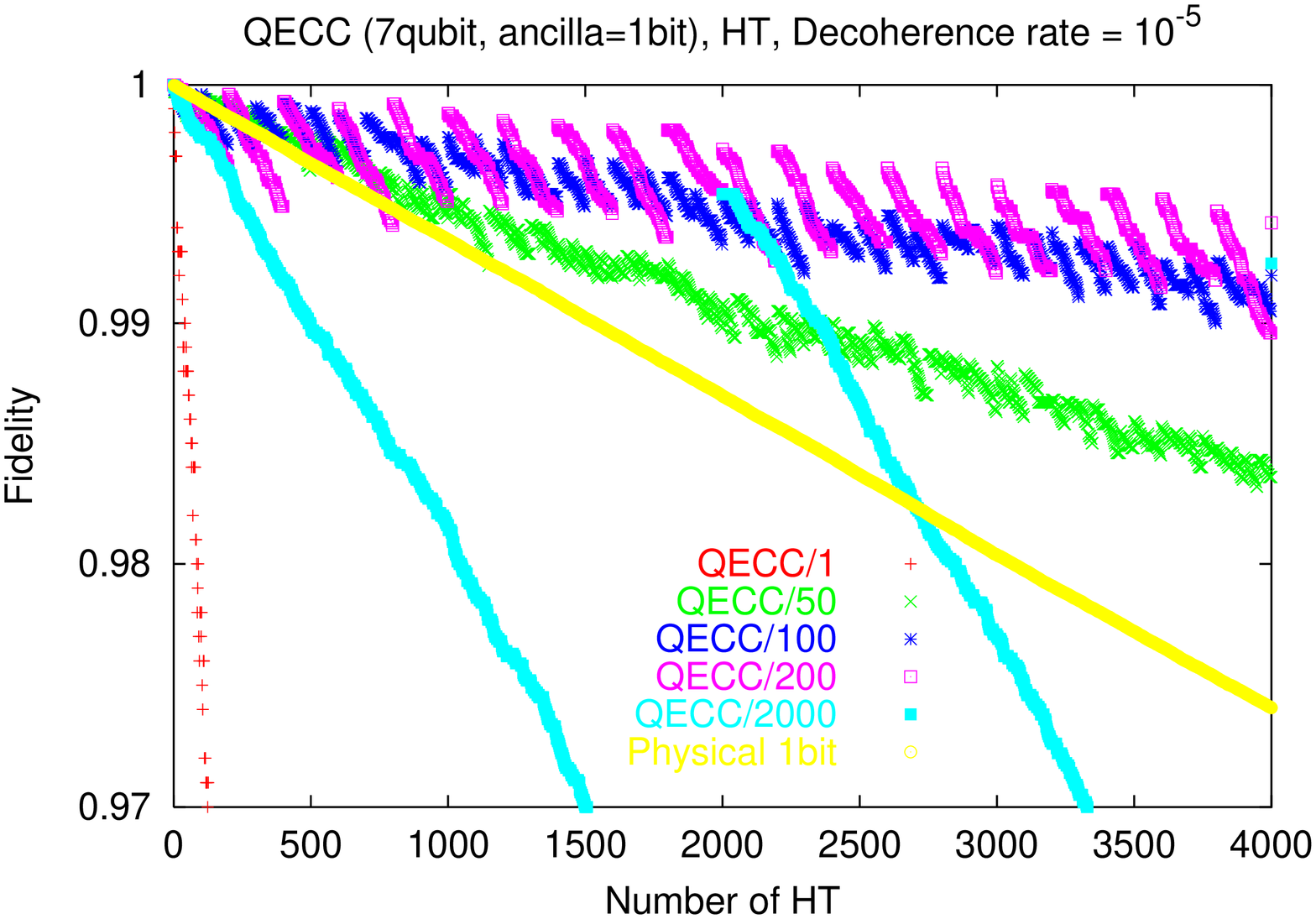}}}
  \resizebox{0.5\textwidth}{!}{\rotatebox{-90}{\includegraphics*{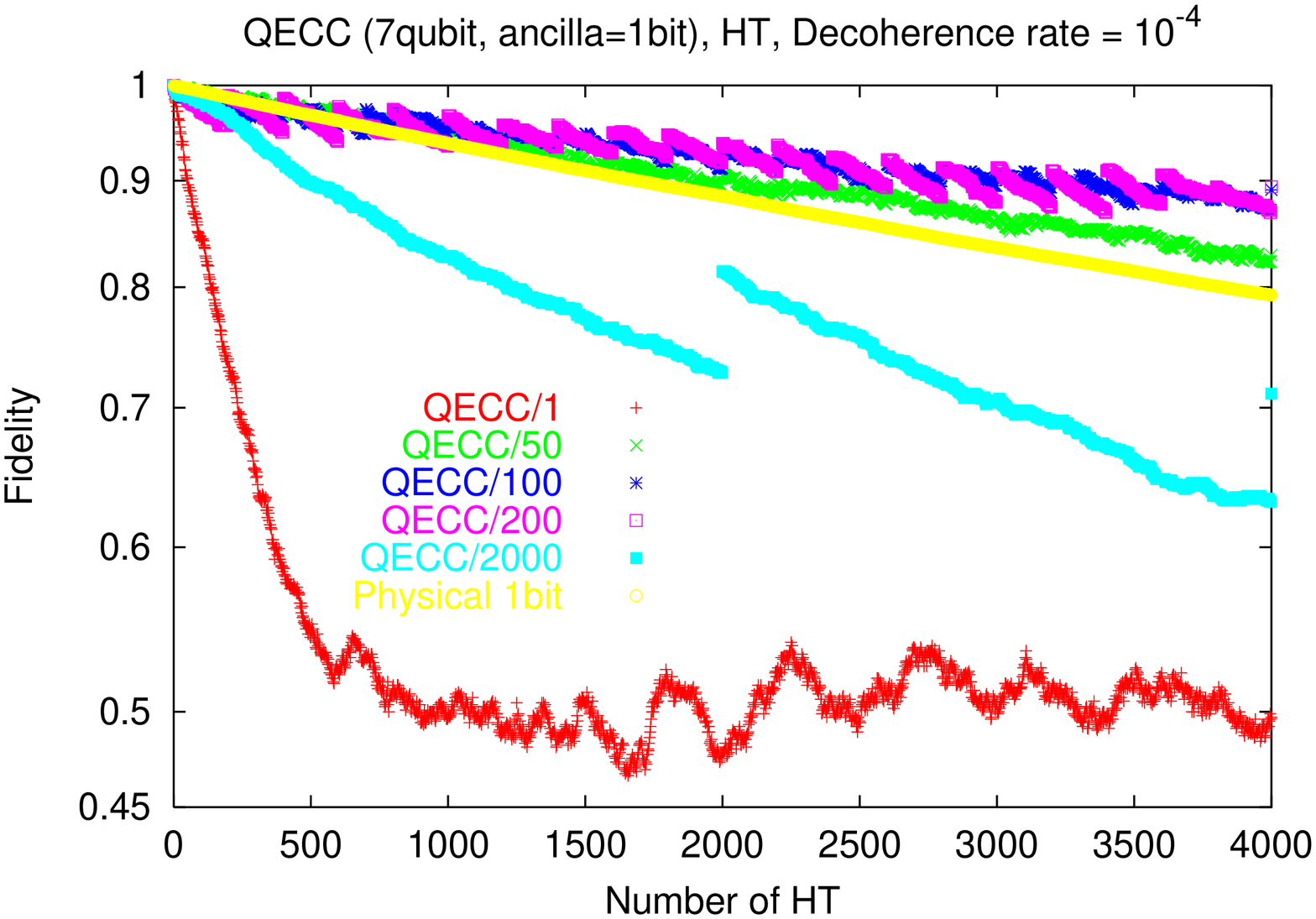}}}\\
  \begin{center}
    \vspace*{-0.7cm}
  \resizebox{0.5\textwidth}{!}{\rotatebox{-90}{\includegraphics*{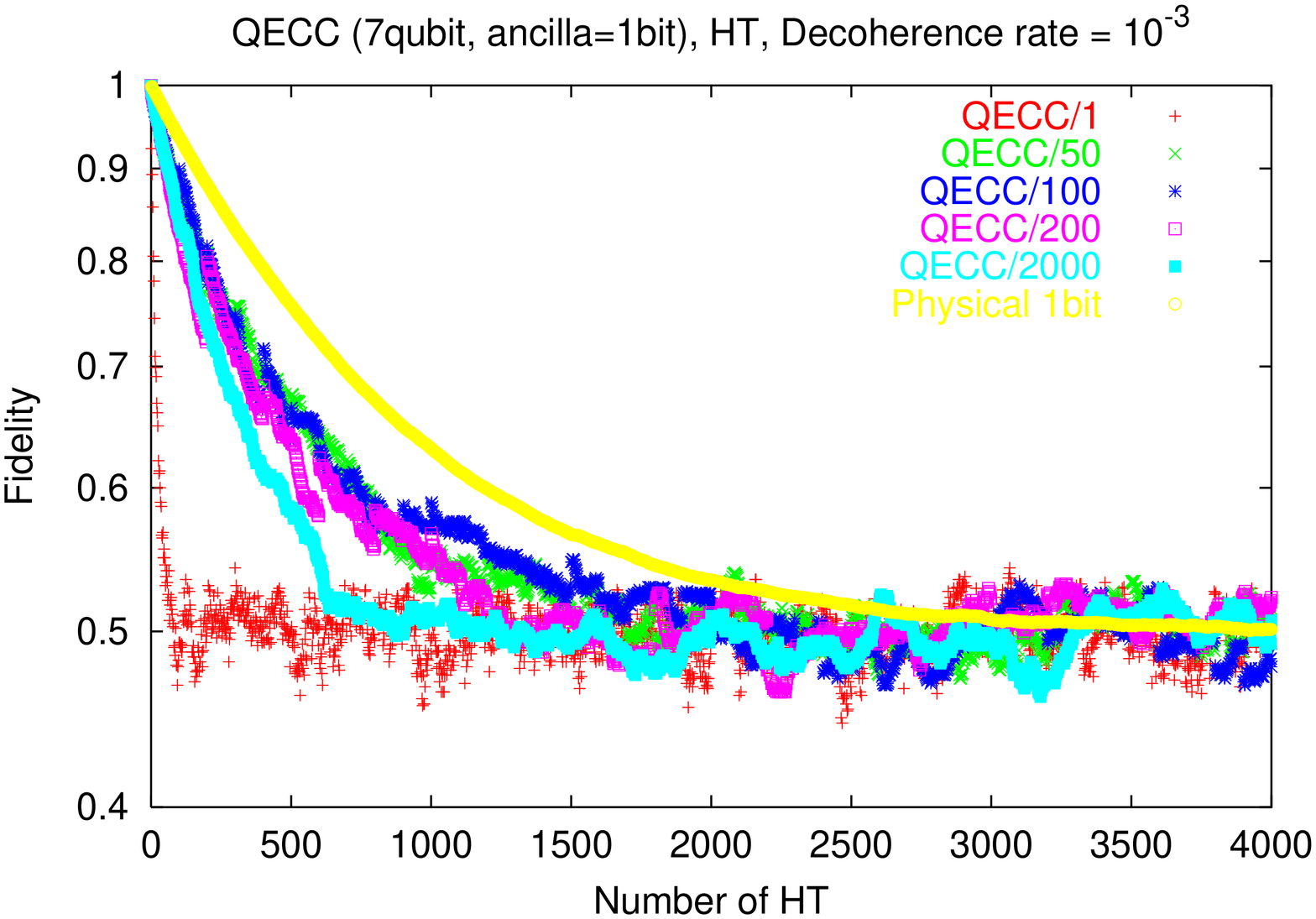}}}    
  \end{center}
  \vspace*{-0.5em}
  \caption{Fidelity with decoherence errors ($10^{-5}$, $10^{-4}$ and
  $10^{-3}$) for the seven qubit code.}
  \vspace*{-0.5em}
\label{fig:7bits}
\end{figure}
Figure \ref{fig:7bits} shows how the frequency of
the error-correction operation affects the fidelity for decoherence 
(probability = $10^{-5}$, $10^{-4}$ and $10^{-3}$) in the seven qubit
code case.
For example, ``QECC/50'' in the Figure means the error-correction
circuit is performed at every 50 main (i.e., Hadamard) gates and
``Physical 1 bit'' means the real (physical) 1bit, hence QEC scheme
cannot be applied to ``Physical 1 bit'' case.

Figure \ref{fig:7bits} shows that fidelity is improved just after the
QEC operation with adequate frequency is performed. However, it also
shows that the fidelity is not high for the circuit that perform
error-correction operations at every $1$ main gates (``QECC/1''). 
When the decoherence probability is $10^{-5}$, we can see that
``QECC/x'' (x=50, 100, 200, 2000) is better than ``Physical 1 qubit''.
When the decoherence probability is $10^{-4}$, the result is slightly
different. ``QECC/2000'' is worse than ``Physical 1 qubit''.
In this situation, we consider that ``QECC/2000'' is not sufficient
and multiple errors occur between QEC circuits, which results in an
uncorrectable state. When the decoherence probability is $10^{-3}$, we
can see that ``Physical 1 qubit'' is better than ``QECC/x'' (x=1, 50,
100, 200, 2000).

It is reasonable to suppose that computation with QEC
scheme is worse than computation without it for any frequency of QEC
operations when decoherence probability is more than $10^{-3}$. These
results show that the appropriate frequency of QEC operations (i.e.,
every $50 \sim 200$ main gate) makes the 7 qubit scheme really
effective when the decoherence probability is not more than $10^{-4}$.

\begin{figure}[hbtp]
  \vspace*{-0.5em}
  \centering
  \resizebox{0.5\textwidth}{!}{\rotatebox{-90}{\includegraphics*{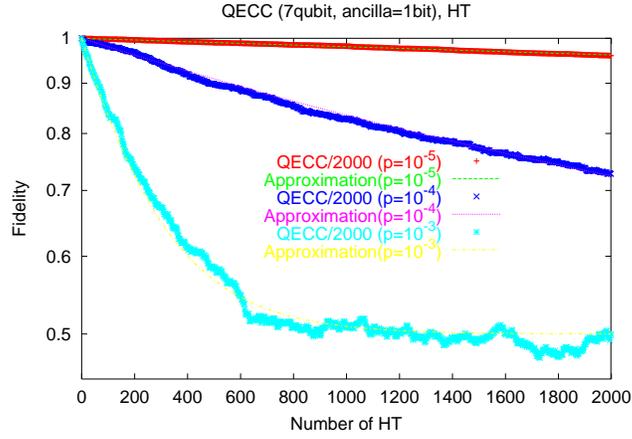}}}
  \vspace*{-0.5em}
  \caption{Comparison between approximation and simulation for the 7
  qubit code (1).}
\label{fig:approx1}
\end{figure}

\paragraph{Analysis}


The ``Physical 1 qubit'' case is theoretically calculated as 
$\frac{1+(1-\frac{4}{3}p)^n}{2}$ ($n$ means the even number of
Hadamard transform and $p$ means the decoherence rate). It
is of course the first-order approximation. We consider two
consecutive Hadamard gates as the unit. Let $1-P$ be the probability that 
the fidelity is decreased at the end of the unit. We assume that there
are at most 1 error in the unit, that is,  before the first Hadamard
gate or just before the second Hadamard gate. $X$ and $Y$ before the
first Hadamard gate degrade the fidelity at the end of the unit, and
$Z$ and $Y$ just before the second Hadamard gate degrade the fidelity
at the end of the unit.  Thus, 
$$1-P = 1-4\left\{\frac{1}{3}p(1-p)\right\} \approx 1 - \frac{4}{3}p$$.
Hence, fidelity is calculated as $\frac{1+(1-\frac{4}{3}p)^n}{2}$.

Let us consider the encoded case. First, to make the analysis simpler,
we deal with the case where syndrome measurement and recovery
operations are not performed. In this case, we apply only the encoded
Hadamard gates. We consider two consecutive encoded Hadamard gates as
the unit. We assume that there are at most one error in the unit.
We generate one error in the unit on purpose and we check which error
really degrades the fidelity at the end of the unit. Through
simulations, we verify that the number of such errors are 12 in
this case. Thus, $$1-P = 1-12\left\{\frac{1}{3}p(1-p)^{13}\right\}
\approx 1 - 4p.$$ 
Therefore, fidelity is calculated as $\frac{1+(1-4p)^n}{2}$. This
approximation is consistent with the simulation results as shown in
Figure \ref{fig:approx1}. We set the number of computation 2000,
Therefore, ``QECC/2000'' means that syndrome measurement and recovery
operations are not performed.

\begin{figure}[htb]
  \vspace*{-0.5em}
  \centering
  \includegraphics*[width=0.6\textwidth]{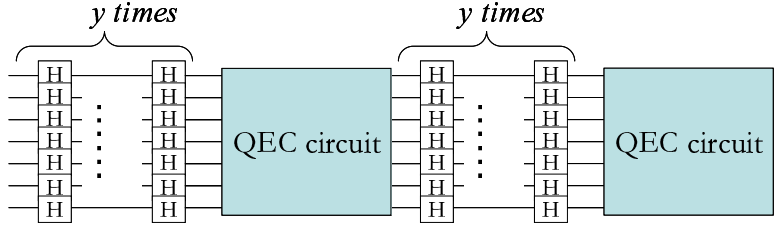}
  \vspace*{-0.5em}
  \caption{The unit for the general case}
\label{fig:g_unit}
\end{figure}

Second, suppose that QEC operations are performed at every $y$ main
(i.e., Hadamard) gate. We consider $y$ consecutive encoded Hadamard
gates, the QEC circuit,  $y$ consecutive encoded Hadamard gates and
the QEC circuit as the unit shown in Figure \ref{fig:g_unit}. 

\begin{figure}[htb]
  \vspace*{-0.5em}
  \centering
  \resizebox{0.5\textwidth}{!}{\rotatebox{-90}{\includegraphics*{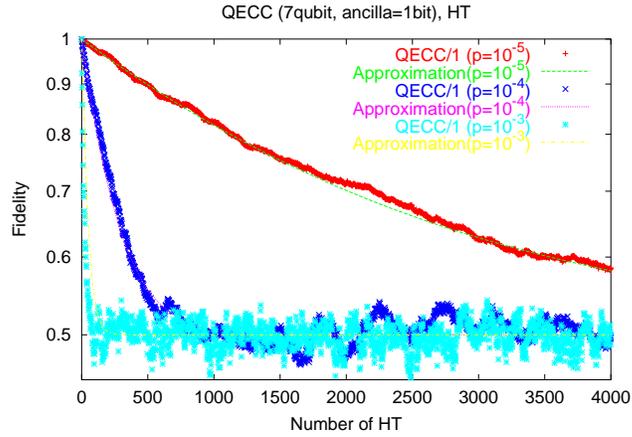}}}
  \vspace*{-0.5em}
  \caption{Comparison between approximation and simulation for the 7
  qubit code (2).}
\label{fig:approx2}
\end{figure}

Again, we assume that there are at most 
one error in the unit.  We generate one error in the unit on purpose
and we check which error really degrades the fidelity at the end of
the unit. Through simulations, we verify that the number of such
errors are 272 and it is independent of $y$.
Thus, 
$$1-P =
1-272\left\{\frac{1}{3}p(1-p)^{6}\cdot(1-p)^{((2\cdot32+2y)-1))\cdot7}\right\}
= 1 - \frac{272}{3}p(1-p)^{447+14y}
\approx 1 - \frac{272}{3}p(1-(447+14y)p).$$
Therefore, fidelity is calculated as 
$\frac{1+(1-P)^{\frac{n}{2 \cdot y}}}{2}$. Of course, this
approximation holds if $p$ is sufficiently small and $y$ is not so
large. Figure \ref{fig:approx2} shows the  fidelity values obtained
from not only simulation results but also  this approximation. We can
see that this approximation is consistent with the simulation results.

\begin{figure}[htbp]
  \vspace*{-0.5em}
  \resizebox{0.5\textwidth}{!}{\rotatebox{-90}{\includegraphics*{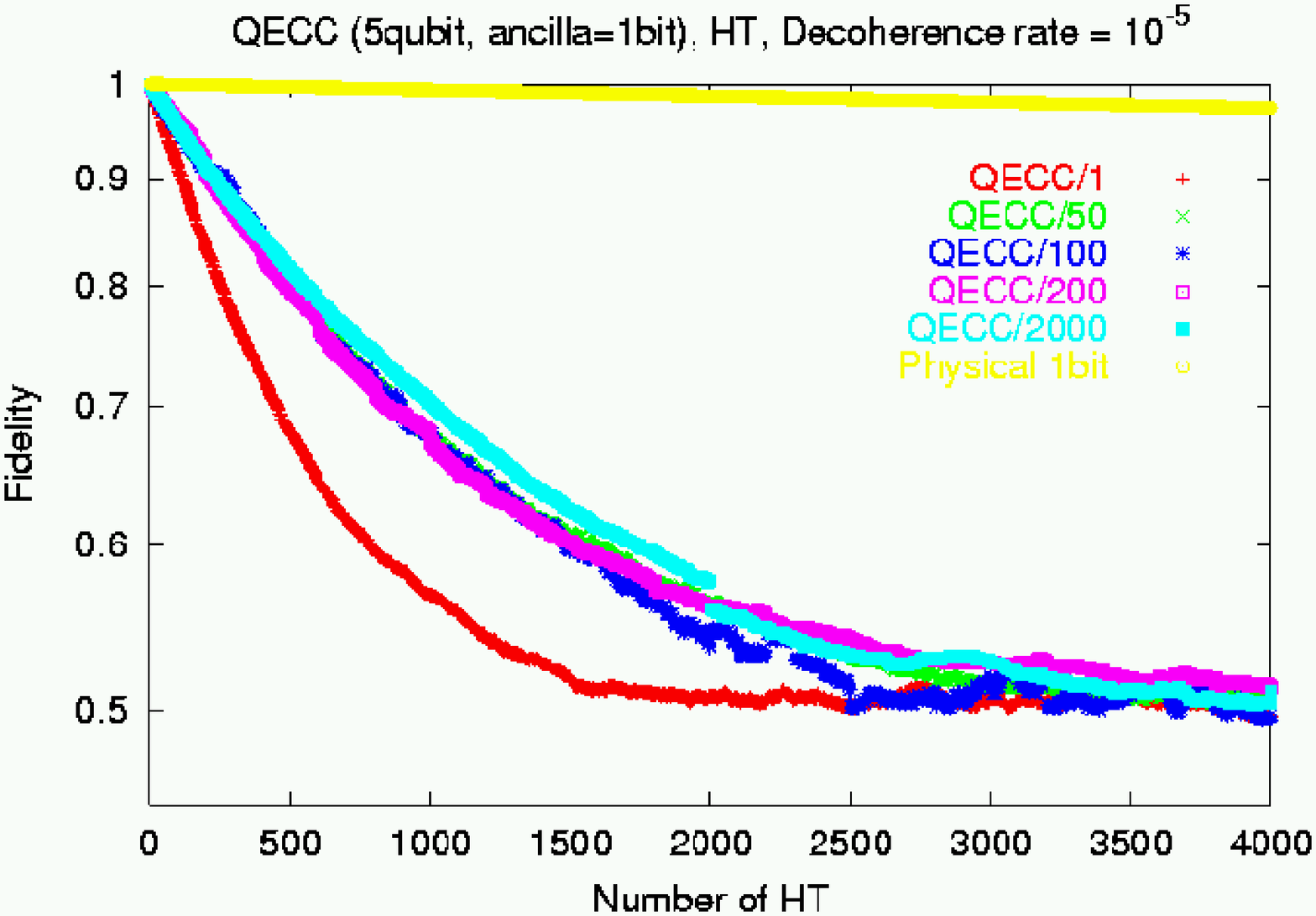}}}
  \resizebox{0.5\textwidth}{!}{\rotatebox{-90}{\includegraphics*{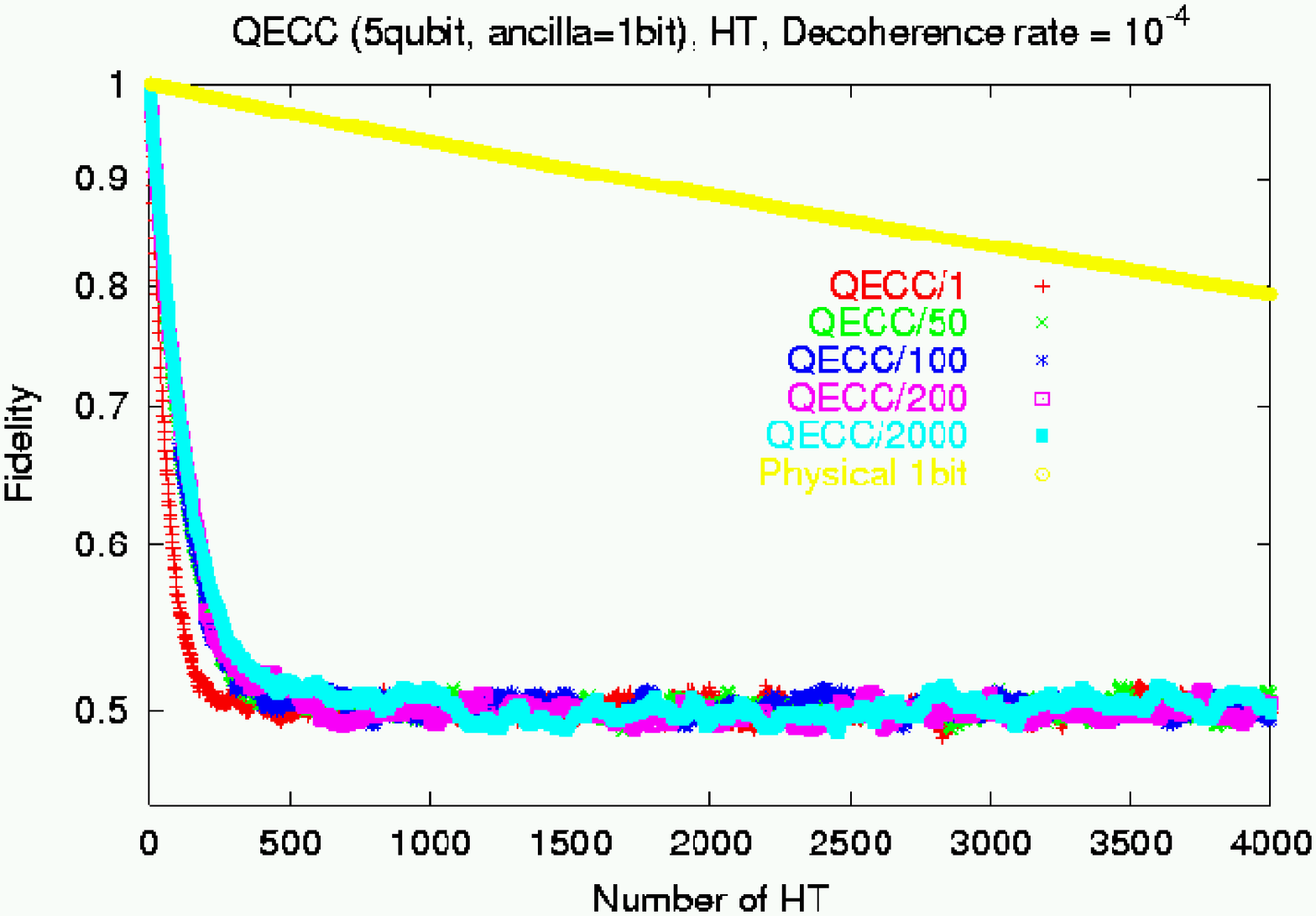}}}\\
  \begin{center}
    \vspace*{-0.5cm}
  \resizebox{0.5\textwidth}{!}{\rotatebox{-90}{\includegraphics*{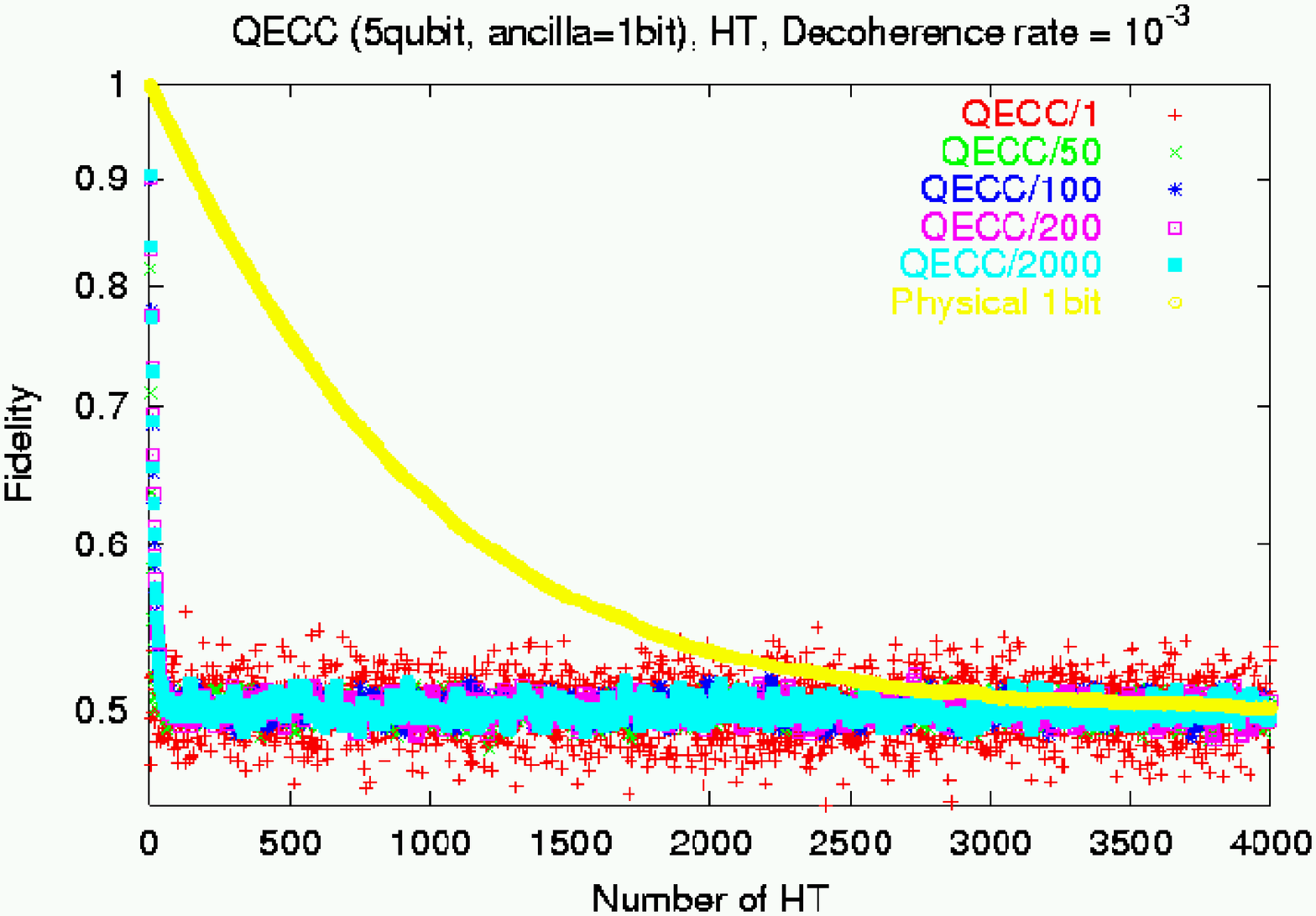}}}    
  \end{center}
  \vspace*{-0.5em}

  \caption{Fidelity with decoherence errors ($10^{-5}$, $10^{-4}$ and
  $10^{-3}$) for the five qubit code.}
  \vspace*{-0.5em}
\label{fig:5bits}
  \vspace*{-0.5em}
  \resizebox{0.5\textwidth}{!}{\rotatebox{-90}{\includegraphics*{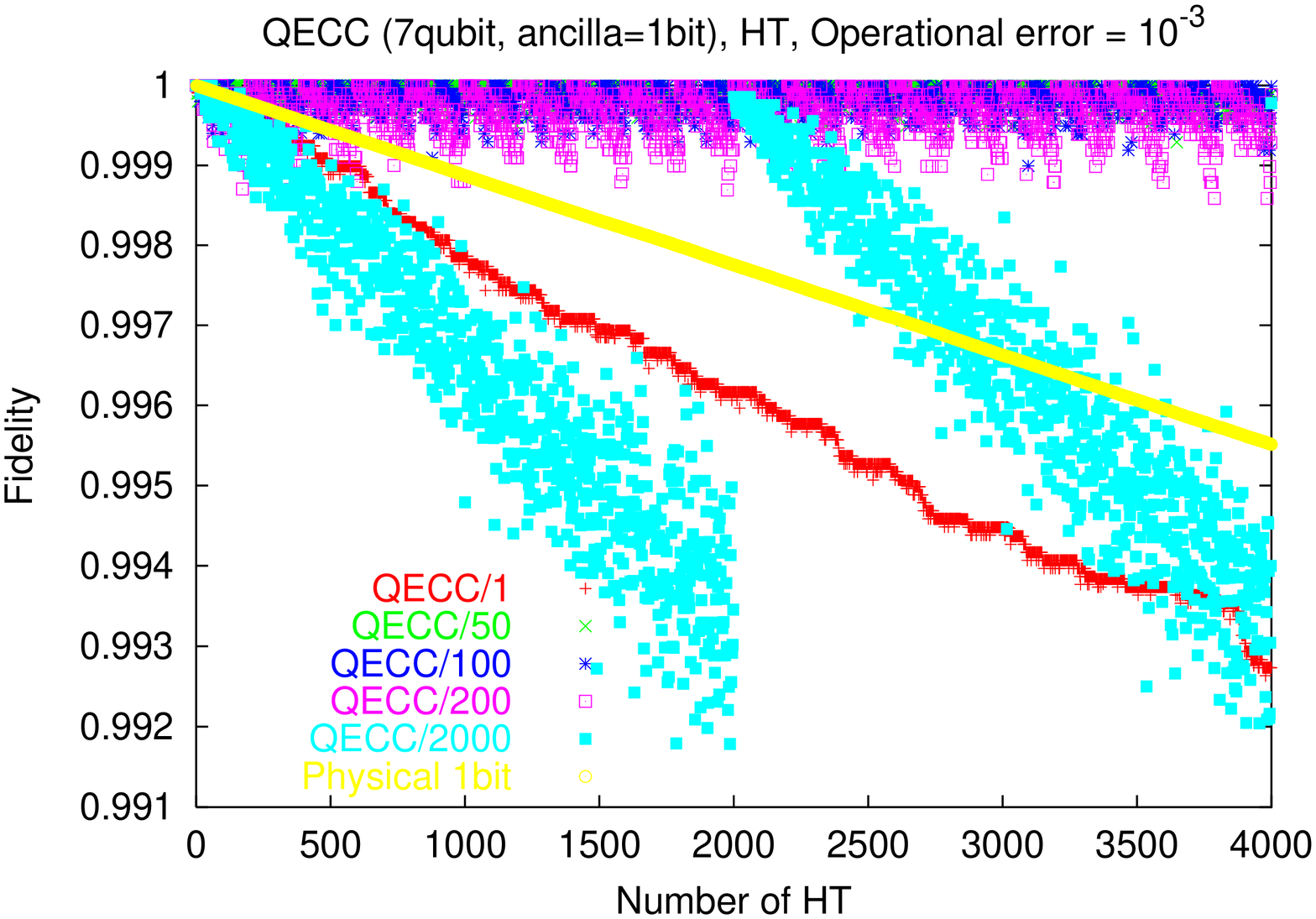}}}
  \resizebox{0.5\textwidth}{!}{\rotatebox{-90}{\includegraphics*{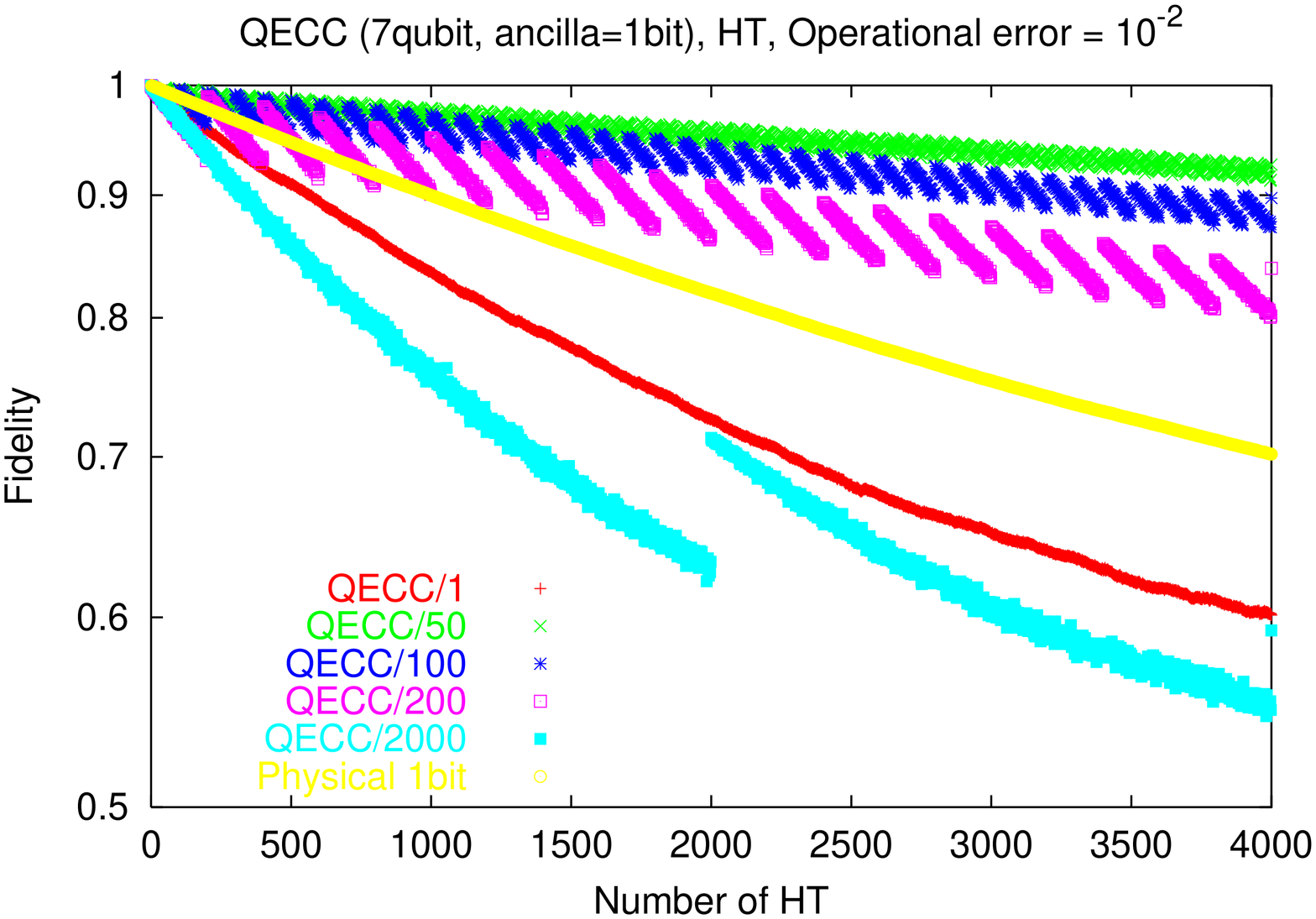}}}\\
  \vspace*{-0.5em}
  \caption{Fidelity with operational errors ($\sigma = 10^{-3}$ and
  $10^{-2}$) for the seven qubit code.}
  \vspace*{-0.5em}
\label{fig:op}
\end{figure}

\paragraph{Five Qubit Code}
Figure \ref{fig:5bits} shows how the frequency of the error-correction
operation affects the fidelity for decoherence 
($10^{-5}$, $10^{-4}$, $10^{-3}$) in the five qubit code case. 

The result clearly shows that computation with error-correction is
much worse than computation without it. The comparison between Figure
\ref{fig:7bits} and Figure \ref{fig:5bits} shows that the fidelity for
the seven qubit code produces better results than that of the five
qubit code. 
The results of nine qubit code is described in the Section
\ref{sec:comp}.

\subsection{Operational Errors}

Figure \ref{fig:op} shows how fidelity is affected by performing QEC
operations in the presence of operational errors. 
We should report that we cannot detect fidelity decrease when the
standard deviation $\sigma$ is no more than $10^{-4}$.

It is found from the results that QEC scheme is effective for
operational errors. This clearly shows that the apparent continuum of 
errors can be corrected by QEC process, that is, by correcting only a
discrete subset of those errors \cite{Shor95:9bit, Steane96}.

This is because Pauli matrices span the space of 2 $\times$ 2 matrices.
Let $|\phi\rangle$ be the state of encoded 1 qubit. To make the
analysis easy, we assume that operational error occurs on the $i$-th
qubit only. As an error operator on $i$-th qubit $E_i$ can be
expanded: 
\[E_i = e_{i0}I+e_{i1}X+e_{i2}Z+e_{i3}Y\]
The quantum state $E_i|\phi\rangle $ can be expressed as
superposition of $|\phi\rangle$, $X|\phi\rangle$, $Z|\phi\rangle$,
$Y|\phi\rangle$. The syndrome measurement process collapse
the state into one of the four states: $|\phi\rangle$,
$X|\phi\rangle$, $Z|\phi\rangle$, $Y|\phi\rangle$.
The recovery process performs the inversion operation and hence the
state recovers. 

\subsection{Both Decoherence and Operational Errors}
\begin{figure}[htb]
  \vspace*{-0.5em}
  \resizebox{0.5\textwidth}{!}{\rotatebox{-90}{\includegraphics*{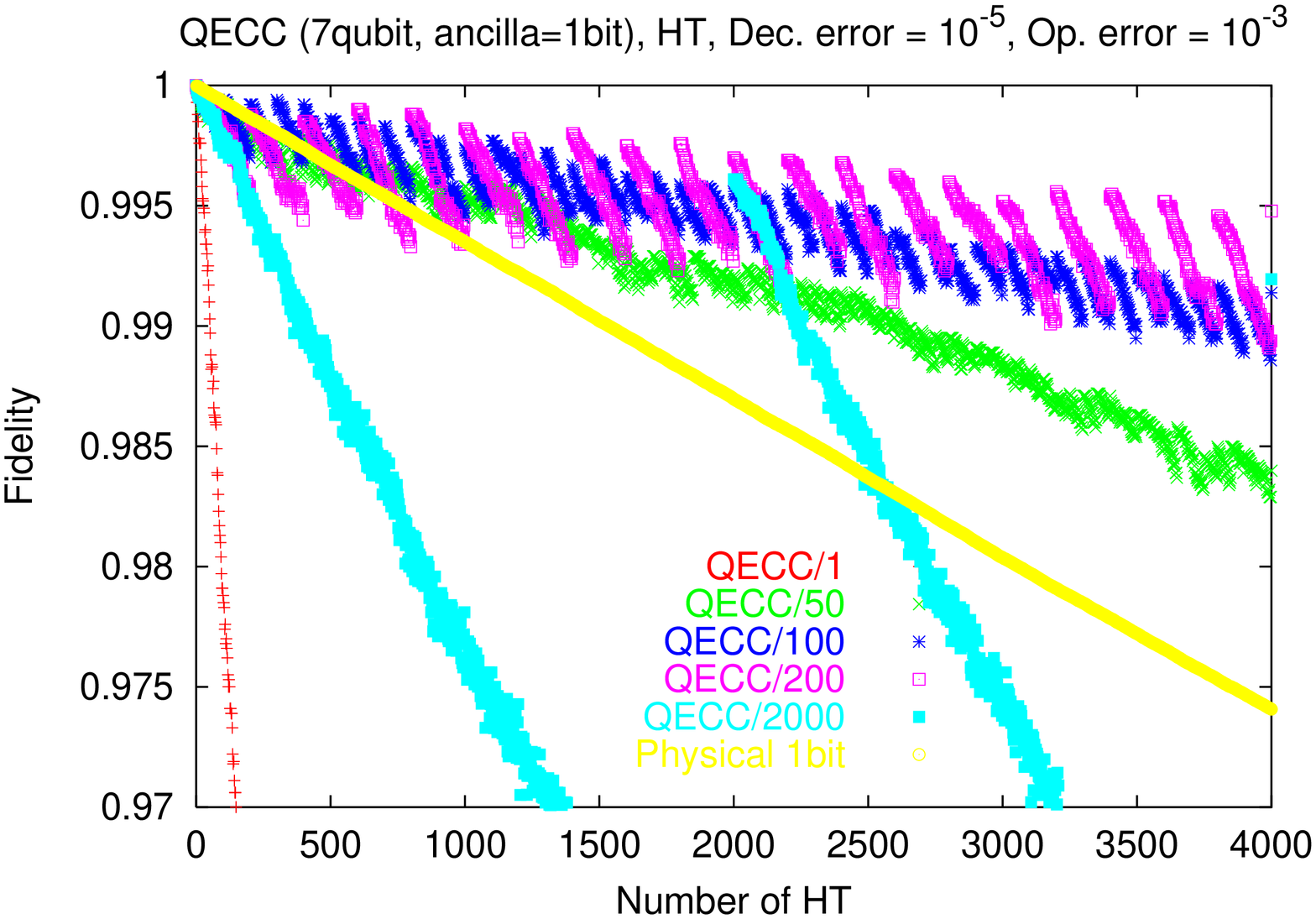}}}
  \resizebox{0.5\textwidth}{!}{\rotatebox{-90}{\includegraphics*{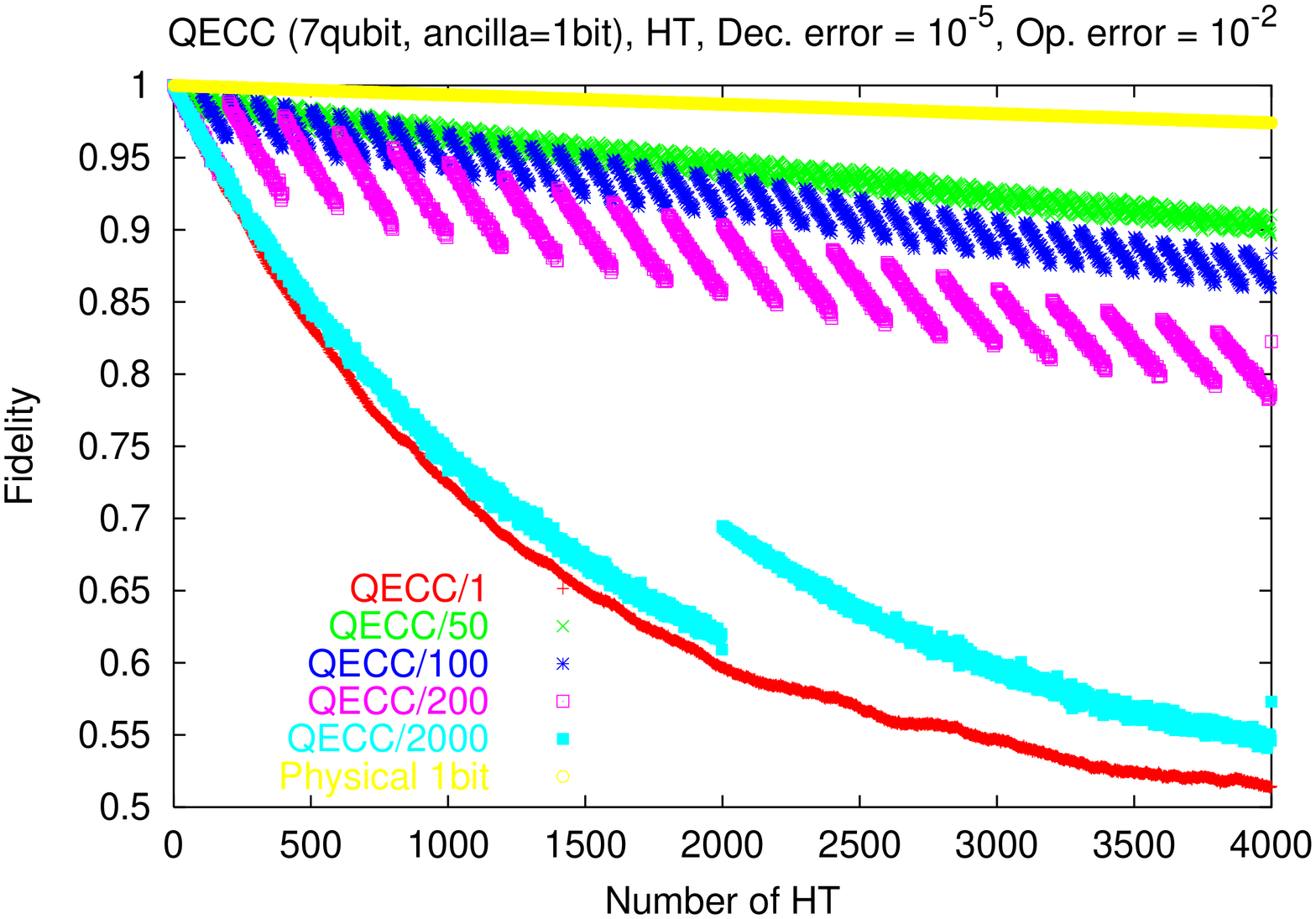}}}\\
  \vspace*{-0.5em}
  \caption{Combined effects for the seven-qubit error-correction
    (($p=10^{-5}$, $\sigma=10^{-3}$) and
    ($p=10^{-5}$,$\sigma=10^{-2}$)).}
  \vspace*{-0.5em}
\label{fig:7d5}
\end{figure}

\begin{figure}[htb]
  \vspace*{-0.5em}
  \resizebox{0.5\textwidth}{!}{\rotatebox{-90}{\includegraphics*{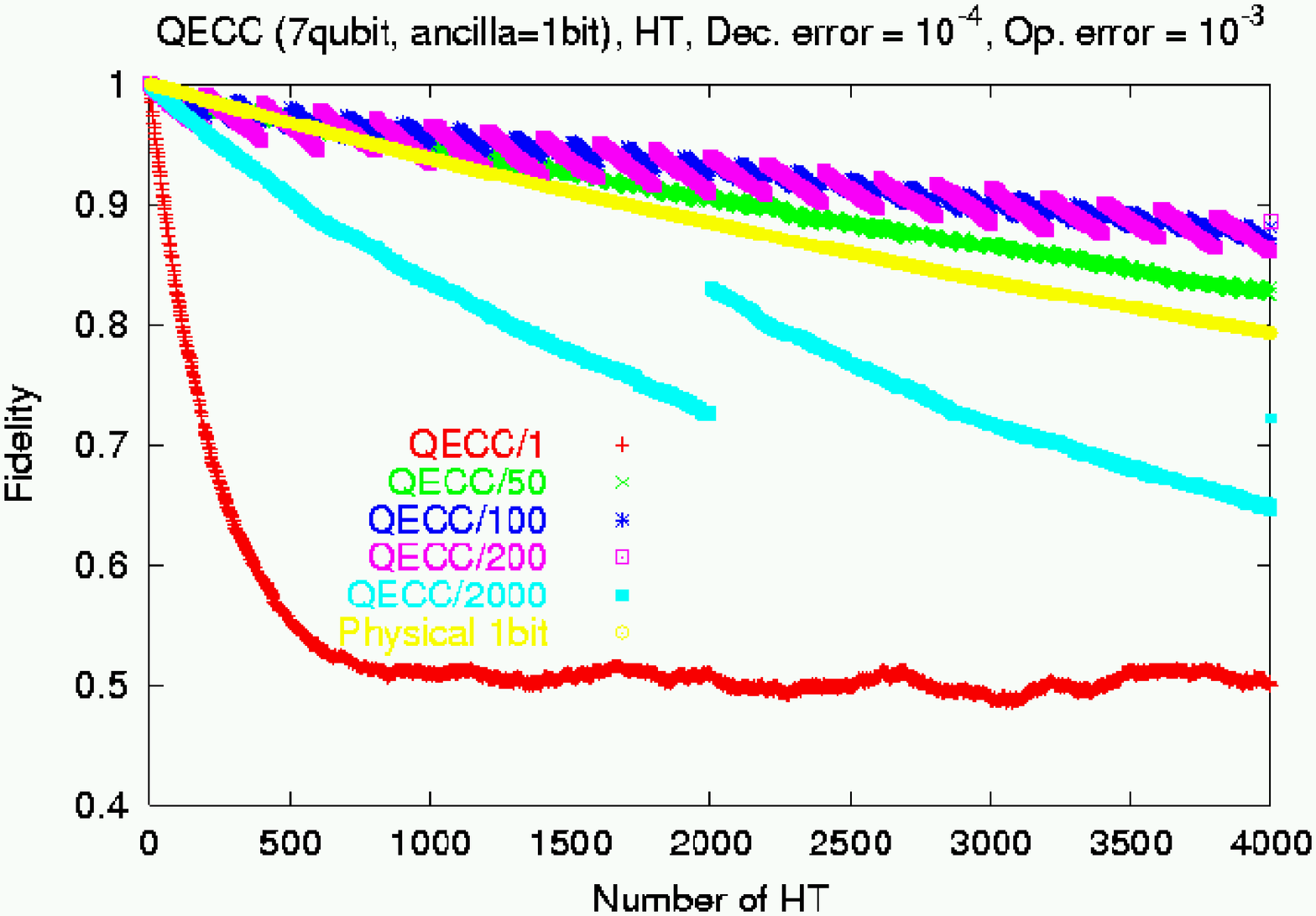}}}
  \resizebox{0.5\textwidth}{!}{\rotatebox{-90}{\includegraphics*{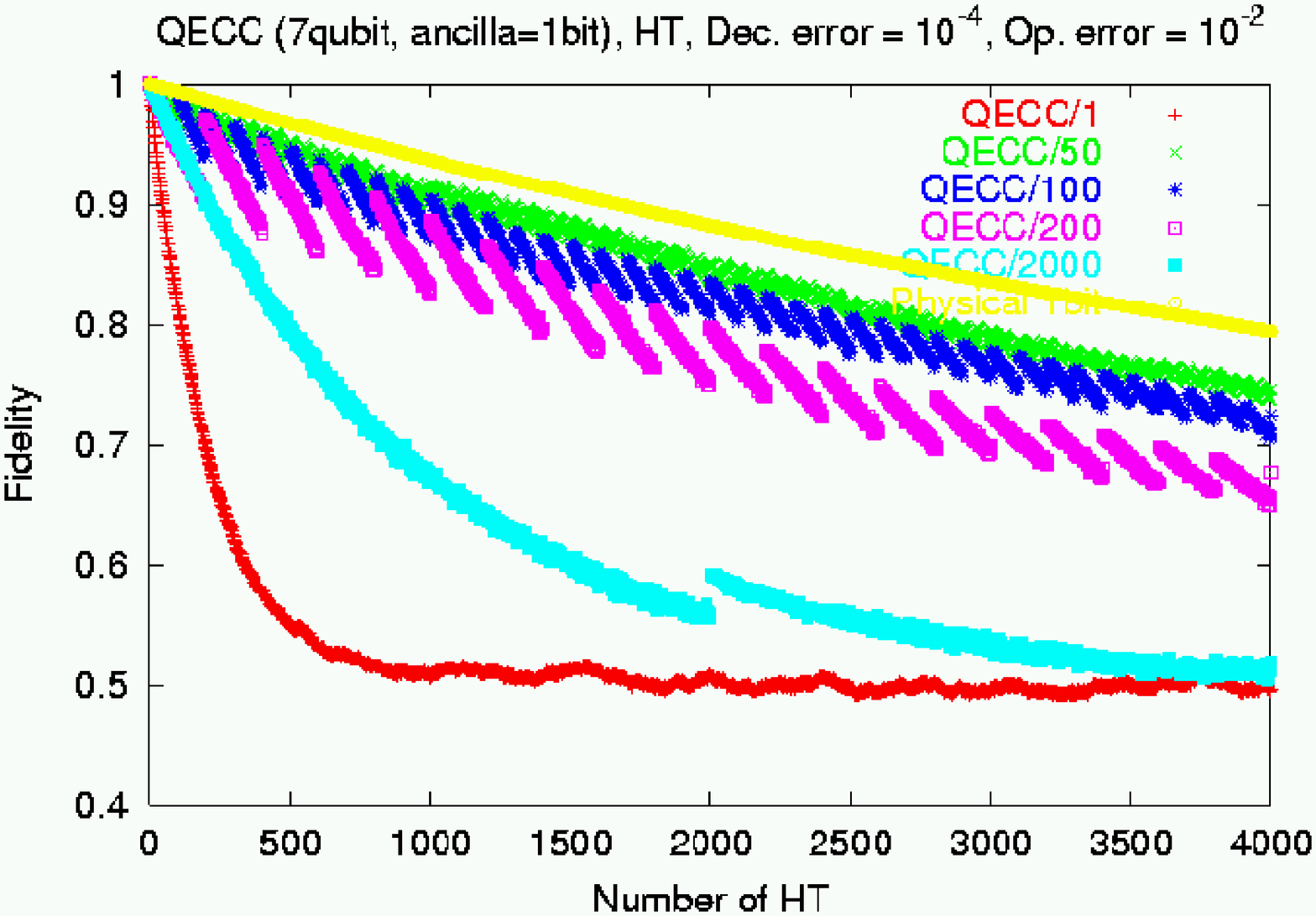}}}\\
  \vspace*{-0.5em}
  \caption{Combined effects for the seven-qubit error-correction
    (($p=10^{-4}$, $\sigma=10^{-3}$) and
    ($p=10^{-4}$,$\sigma=10^{-2}$)).}
  \vspace*{-0.5em}
\label{fig:7d4}
\end{figure}

\begin{figure}[htb]
  \resizebox{0.5\textwidth}{!}{\rotatebox{-90}{\includegraphics*{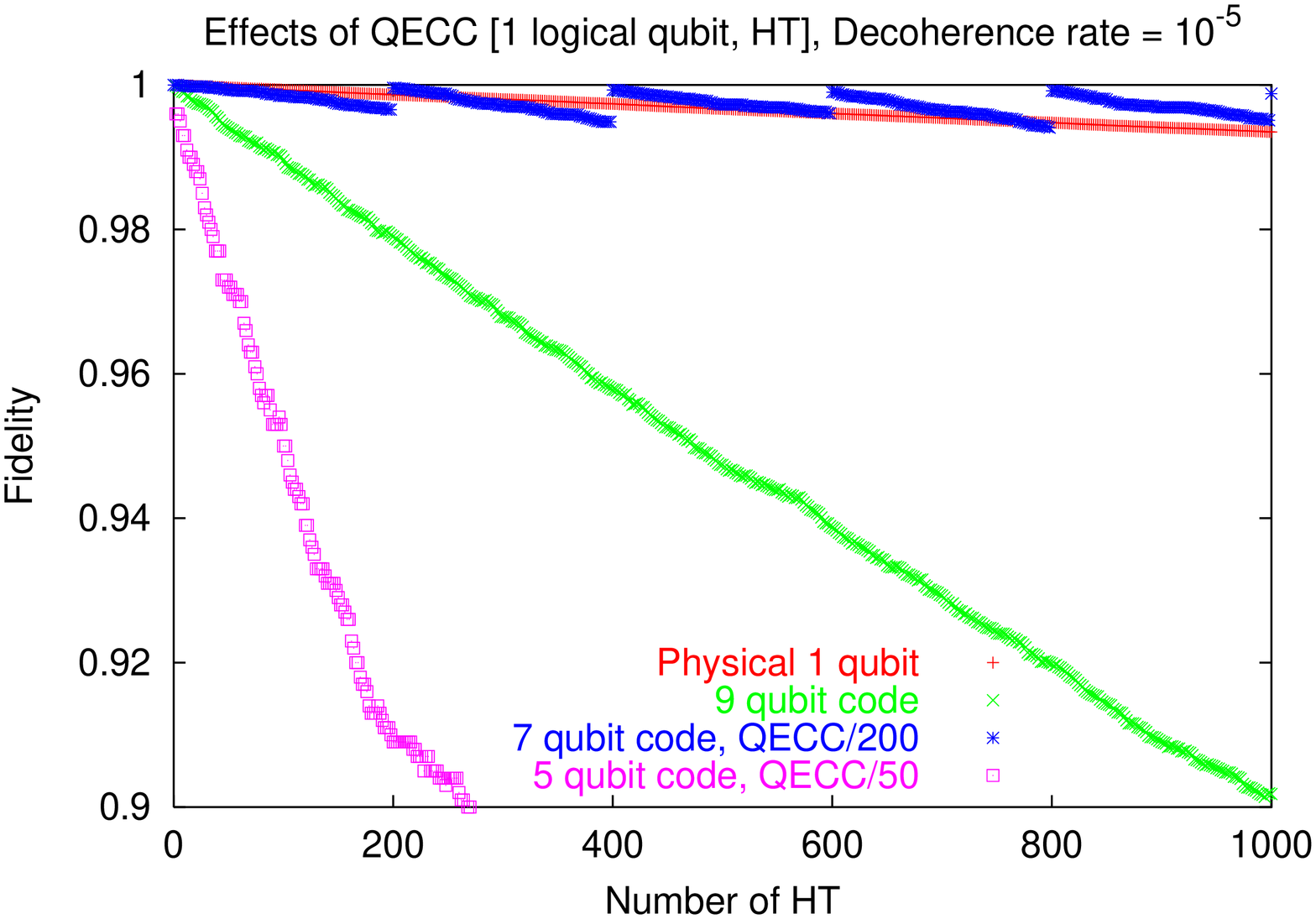}}}
  \resizebox{0.5\textwidth}{!}{\rotatebox{-90}{\includegraphics*{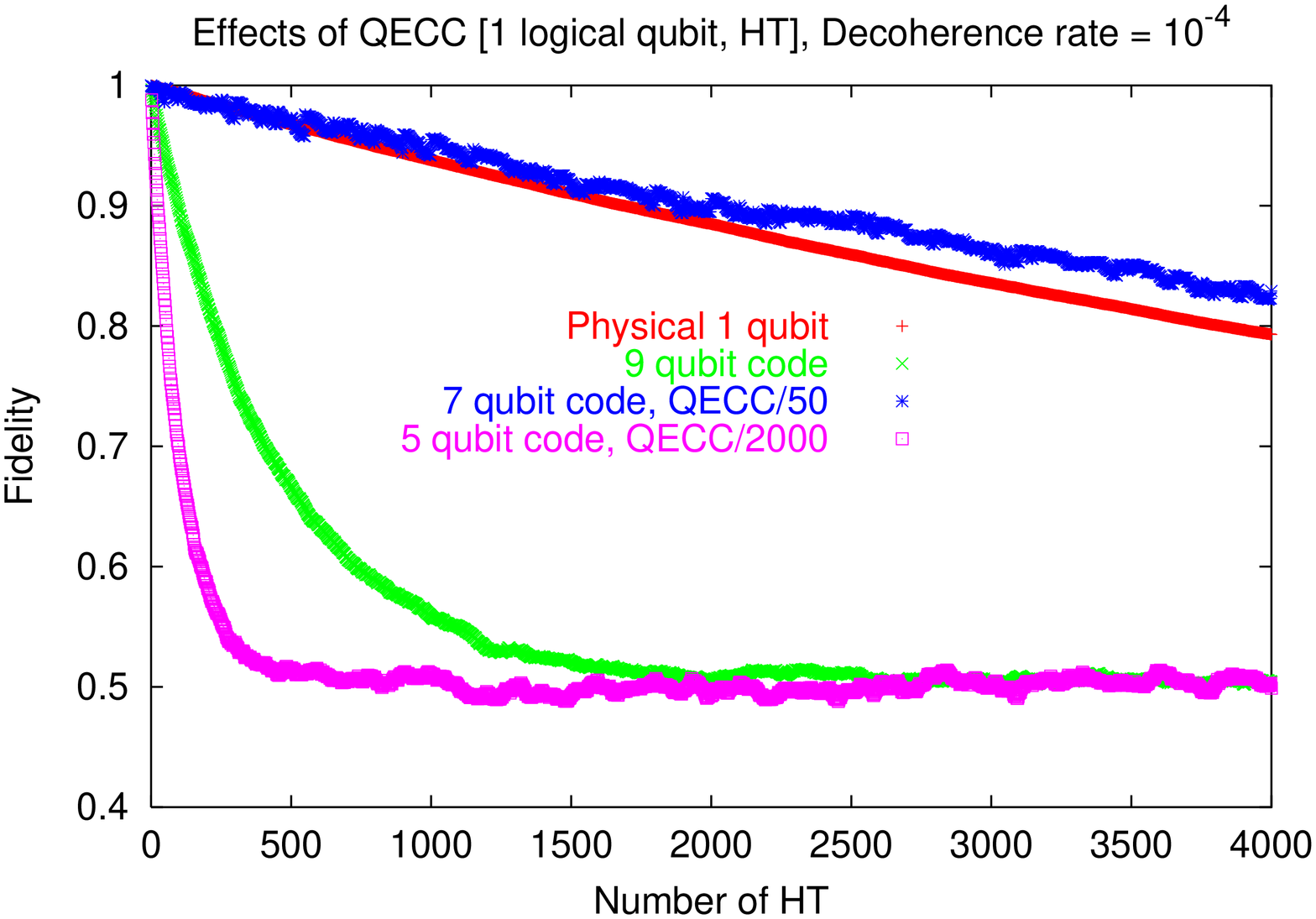}}}\\
  \begin{center}
    \resizebox{0.5\textwidth}{!}{\rotatebox{-90}{\includegraphics*{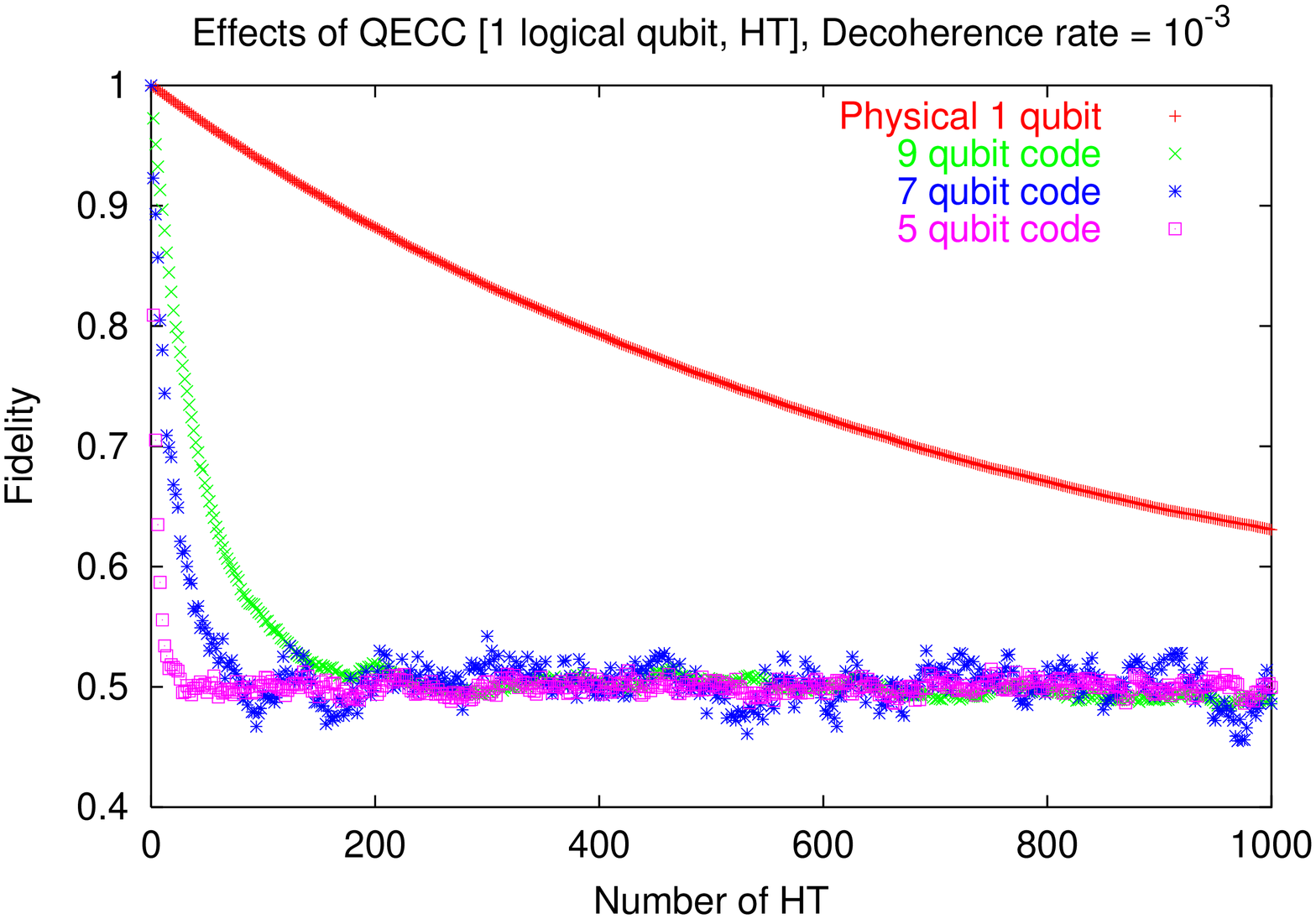}}}    
  \end{center}
  \caption{Fidelity with decoherence errors ($10^{-5}$,
    $10^{-4}$, $10^{-3}$) for nine qubit code, seven qubit code and 
    five qubit code.}
\label{fig:1G}
\vspace*{-0.5cm}
\end{figure}

Now, we take account of not only decoherence errors but also
operational errors. It is difficult to analyze theoretically the
effects of QECC when both errors exist. Therefore, numerical
simulations are needed.

We use the seven qubit code because we have confirmed that it is
effective at least for each error. 
Figure \ref{fig:7d5} and \ref{fig:7d4} show the combined effects of
both errors. In this experiment, ``Physical 1 qubit'' represents the
fidelity of physical 1 qubit with only decoherence errors. 
The combined effect of two factors may be worse than each factor
alone, that is to say, the effect seems to be the product of each
factor. As these Figures indicate, computation with QEC scheme is
worse than computation without QEC scheme when the standard deviation
of operational errors is $10^{-2}$. Therefore, it is reasonable to
suppose that the QEC scheme using the seven qubit code is still
effective even if both errors exist, as long as the decoherence rate
is not more than $10^{-4}$ and the standard deviation of operational
errors is not more than $10^{-3}$ and the frequency of QEC operations
are appropriate (i.e., at every $50 \sim 2000$ main gate).

\subsection{Code Comparison}
\label{sec:comp}
Figure \ref{fig:1G} shows the fidelity with decoherence errors
($10^{-5}$, $10^{-4}$ and $10^{-3}$) for the QEC circuit of nine qubit
code described in Section 2.  As stated above, the error-correction
operation for nine qubit code scheme must be performed at every 1 main
gate.  For comparison, Figure \ref{fig:1G} also shows the best
fidelity for five qubit code and that for seven qubit code.

From the Figure  \ref{fig:1G}, we can see that seven qubit code is
best. As stated in Section \ref{sec:sm}, the five qubit code scheme
and the nine qubit code scheme always require the encoding and
decoding process per 1 main computation step. From the Table
\ref{tab:comp}, we find that the five qubit scheme requires at least
21 steps to perform 1 main computation step and the nine qubit scheme
requires at least 13 steps, whereas the seven qubit code scheme
requires at least 1 step. This difference leads to the fidelity
difference shown in Figure \ref{fig:1G}.

We can also find that the fidelity for the nine qubit code is higher
than that for the five qubit code. We consider this is partly because
the QEC circuit of the nine qubit code scheme is simpler and hence
induces less errors.  When the decoherence rate  is $10^{-3}$, the
fidelity for the nine qubit code is best during the first 200 times
computation, which is contrary to our intention.

\section{Related Work}
There are several numerical studies paying attention to the realistic
case in which errors occur during the error-correction operations.
In Ref\cite{ObenlandPhd}, the seven qubit code is simulated in the
Obenland's simulator based on an actual physical experimental
realization.  Its error model is the same as our model.

In Ref\cite{SALAS02}, the seven qubit code is used and
the ability of Steane's and Shor's ancilla are tested by Monte Carlo
simulations. In this paper, on the other hand, not only the seven
qubit code but also the five qubit code and the nine qubit code are
dealt with and their effects are simulated and compared.

In Ref\cite{Steane97:SIM, Steane02:SIM}, more various CSS codes (such
as Golay code $[[23, 1, 7]]$) are dealt with. In these studies, error
models are different from ours. They use gate errors instead of
operational errors. Every gate is modeled by a failure followed by a
perfect operation of the gate. The failure for a single-qubit gate is
the same as the decoherence error and the failure for two-qubit gate is
modeled as follows. With probability $1-q$ there are no changes and
with equal probabilities $\frac{q}{15}$, 15 possible one or two qubit
errors occur $\{IX,IY,IZ,XI,XX,XY,XZ,YI,YX,YY,YZ,ZI,ZX,ZY,ZZ\}$. 
They keep track of not the quantum states but the error operators ($X,
Y, Z$).

\section{Summary}
It is necessary to investigate the effects of quantum error-correction
schemes in the realistic case. We have investigated the effects of QEC
scheme by simulations. We confirmed that the seven qubit code scheme is
really effective when decoherence rate is not more than $10^{-4}$ and
the standard deviation of operational error is not more than $10^{-3}$
and the error-correction process is performed at every 50 $\sim$ 200
main gate even if both errors exist. We also have confirmed that
transversal operations are really important. However, we 
have found that we cannot say that the four qubit ancilla system is
better than the one qubit ancilla system.

For the future work, we will implement transversal gate for the five
qubit code scheme and the nine qubit code scheme. We will investigate
other CSS codes. We will adopt more complex computation as main
computation, such as QFT and Grover circuit. Further investigations
are required for fault-tolerant measurements with which we are not
concerned in this paper and the speed of gate operations should be
considered.

\section*{Acknowledgment}
We would like to thank Dr. Mitsuru Hamada for valuable comments and
advice for QECC.
\bibliographystyle{plain}
\bibliography{Paper}
\end{document}